\documentclass[reprint,
superscriptaddress,
 amsmath,amssymb,
 aps,prl,]{revtex4-1}
\usepackage{color}
\usepackage{graphicx}
\usepackage{dcolumn}
\usepackage{bm}

\begin{document}

\preprint{APS/123-QED}

\title{Route to Achieving Giant Magnetoelectric Coupling in BaTiO$_3$/Sr$_2$CoO$_3$F \\
Perovskite Heterostructures}

\author{Indukuru Ramesh Reddy}
\affiliation{Department of Physics, National Institute of Technology Karnataka, Srinivasnagar, Surathkal, Mangalore-575025 India}
\author{Peter M.\ Oppeneer}
\email{peter.oppeneer@physics.uu.se}
\affiliation{Department of Physics and Astronomy, Uppsala University, Box 516, S-75120 Uppsala, Sweden}
\author{Kartick Tarafder}
\email{karticktarafder@gmail.com}
\affiliation{Department of Physics, National Institute of Technology Karnataka, Srinivasnagar, Surathkal, Mangalore-575025  India}

\date{\today}

\begin{abstract}
\noindent
Polarization induced spin  
switching of atoms in magnetic materials opens for possibilities to design and develop advanced spintronic devices, in particular, storage devices where the magnetic state can be controlled by an electric field. We employ density-functional theory calculations to study the magnetic properties of a perovskite strontium cobalt oxyfluoride Sr$_2$CoO$_3$F (SCOF) in a hybrid perovskite heterostructure, where SCOF is sandwiched between two ferroelectic BaTiO$_3$ (BTO) layers. Our calculations show that the spin state of the central Co atom in SCOF can be controlled by altering the polarization direction of the BTO, specifically, to switch from high-spin state to low-spin state by changing the relative orientation of the ferroelectric polarization of BTO with respect to SCOF, leading to an unexpected, giant magnetoelectric coupling, $\alpha_s \approx 21 \times 10 ^{-10}$ Gcm$^2$/V.
\end{abstract}
\maketitle


Achieving high-speed switching, low power consumption, and high operational stability simultaneously in conventional magnetic storage media is a current challenge. Multiferroic materials with strong magnetoelectric (ME) coupling  may provide a solution to many of these challenges \citep{Eerenstein2006,Trassin2015}. A strong ME coupling in a material enables to control the magnetic properties by means of an electric field and \textit{vice versa}, a functionality that is  promising for a wide range of potential applications such as sensors, high-density nonvolatile memory devices and ME actuators \citep{Scott2012,Hu2011,Hu2012,Wang2014}. In transition metals, the partially filled $d$ states are responsible for the occurrence of magnetism. On the other hand, the empty $d$ states may exhibit ferroelectricity. Consequently, the coexistence of ferroelectricity and magnetism at room temperature in an intrinsic material is very difficult to observe \citep{Scott2007}. Suitable alternatives could be multiferroic heterostructures and composites that comprise both magnetic materials and ferroelectric oxides,  as  was reported in recent years \citep{Vaz2010,Scott2013,Lu2015,Ma2011,Stroppa2011,Fechner2012,DiSante2013,Stroppa2013}.
\par In metal-oxide and oxyhalide based perovskites, the covalent interactions between the metal center and its oxide/halide ligand are at the heart of the trade-off 
between charge, spin, and orbital degrees of freedom, leading to interesting emergent physical properties, such as high-spin (HS) to intermediate/low-spin (IS/LS) state transition, metal-insulator transition, and magnetoresistance phenomenon, to name a few \citep{Moritomo2000,Briceno2018}. The recently synthesized  perovskite layered structure of strontium cobalt oxyfluoride Sr$_2$CoO$_3$F (SCOF) represents an interesting coordination framework, wherein the coordination of the metal Co atom can be switched from octahedral to square pyramidal shape by an external stimulation. In the ground state SCOF behaves as an  antiferromagnetic insulator with  relatively high N\'eel temperature, $T_N=323$~K \citep{Oxyfluorides2012}.  A pressure driven spin-crossover has recently been observed in layered SCOF.  Tsujimoto \textit{et al.}\ \citep{Tsujimoto2016} showed that the spin state of the Co$^{3+}$ ion switches from HS to LS configuration with an applied pressure of 12 GPa. 
The main mechanism behind the  spin state  switching of the Co atom with applied pressure is a shrinking of the Co--F bond length, in which the coordination of the Co atom changes from square pyramidal to octahedral, yet without structural phase transition of SCOF \citep{Tsujimoto2016}.
Ou \textit{et al.}\ \citep{Ou2016}  showed that the  spin-state transition is still possible at a relatively low applied pressure. Their first-principles calculation showed that the spin state switches from HS to IS with an applied pressure $\sim 6$ GPa. They also showed that the relative position of the fluorine atom plays a crucial role in determining the spin state of the Co atom. 
A different, so far unexplored route to perturb the local structure of SCOF and switch its spin state could be to place the material in a charged environment. 

With this in view, we study here the magnetic properties of a SCOF layer under the influence of an induced polarization, created in 
a multiferroic heterostructure that consists of SCOF between two ferroelectric  BaTiO$_3$ perovskite layers. The key idea is to design seamlessly matching perovskite heterostructures and to use the internal polarization of BTO to drive the spin-crossover in SCOF.
Using first-principles 
calculations to investigate the magnetic properties of the designed heterostructures 
we show that the spin state of the Co$^{3+}$ atom can be efficiently controlled by the BTO polarization. 
 Computing the magnetoelectric coupling strength of the BTO-SCOF-BTO multiferroic heterostructure we find that it exhibits a giant ME coefficient
 that is larger than current record values.


To investigate the magnetic properties of BTO-SCOF-BTO heterostructures we employ density-functional theory (DFT) calculations using the 
VASP code \citep{Kresse1996}. 
For the exchange-correlation functional we used the generalized-gradient approximation (GGA) in the  Perdew-Burke-Ernzerhof (PBE) \citep{Perdew1996} parametrization. We implemented DFT+$U$ technique to capture the strong electron-electron correlation which exists in the partially filled 3$d$ shell and is missing in the conventional GGA \citep{Dudarev1998}. This has been proven to be an accurate technique to achieve the precise spin state of molecules and low dimensional magnetic systems  \citep{Sarkar2011,Tarafder2012,Maldonado2013}. The on-site Coulomb and exchange parameters $U$ and $J$ were chosen to be 5.0 eV and 1.0 eV, respectively, for the Co atom.  For further computational details, see Ref.\  \cite{calcs}.

\begin{figure}[t]
\includegraphics[width=6cm]{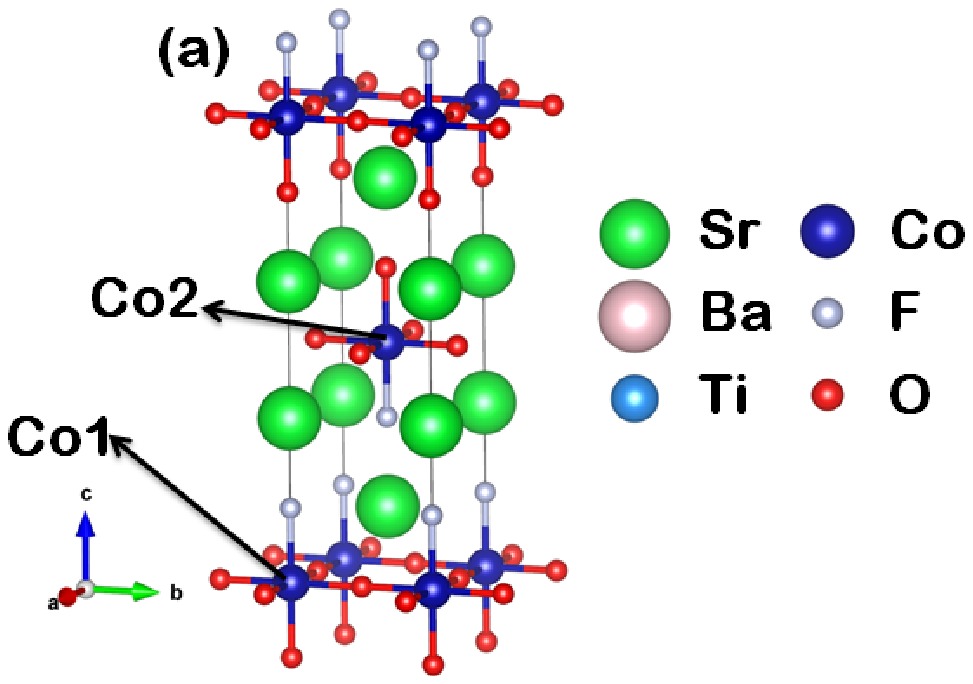}
\includegraphics[width=8cm]{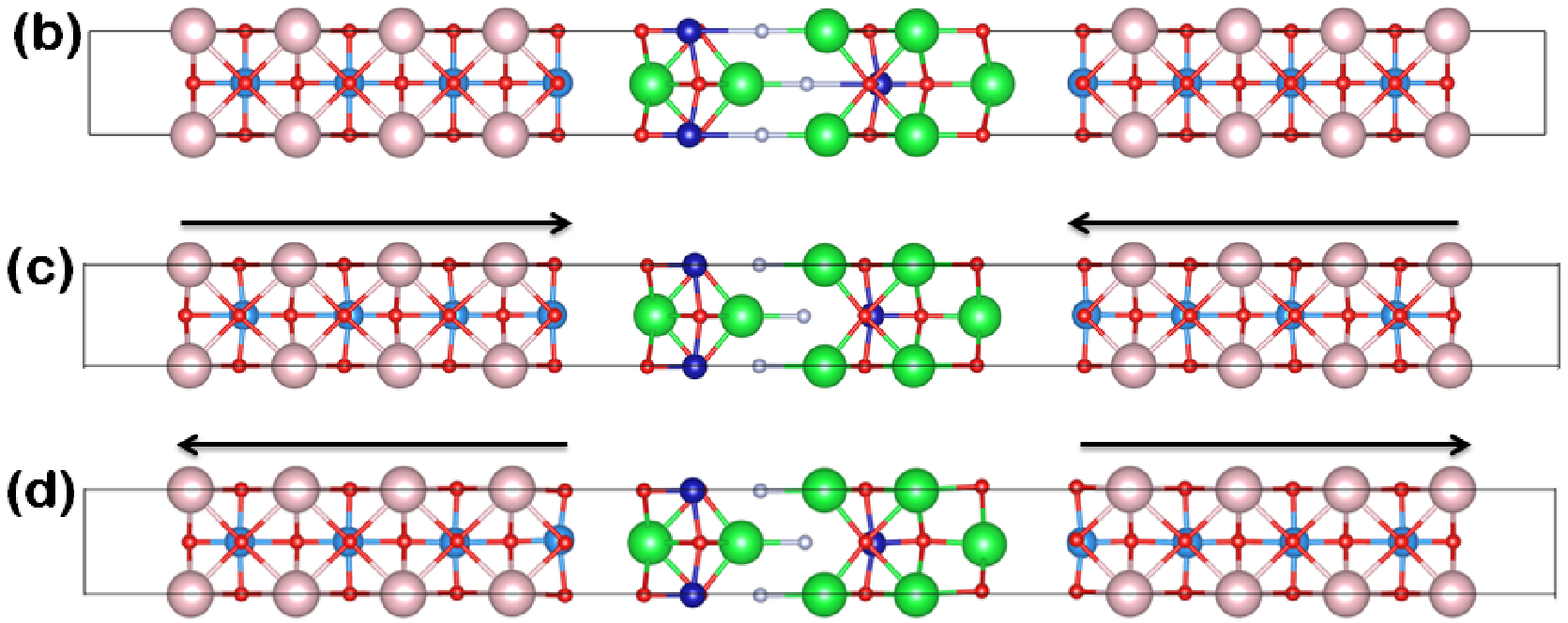}
\caption{\label{fig1} Optimized structures of bulk SCOF and three types of BTO-SCOF-BTO multilayer configurations stacked along [001] direction.  (a) Bulk unit cell of SCOF, (b) nonpolar system (NP), where the BTO composition is pseudo-cubic with no polarization, (c) polar system (P1), the polarization direction of BTO is toward SCOF, and (d) polar system (P2) where the polarization is away from SCOF. Arrows in (c) and (d) indicate the polarization orientation.}
\end{figure}

We carried out the investigation in two steps. To start with, we studied the magnetic properties of bulk SCOF in its optimized geometry. The initial lattice parameters and coordinates of cubic and tetragonal BTO were obtained from previously reported studies \citep{Wang2010,Megaw1962}. Subsequently, a full structural optimization of all bulk structures was carried out. The polarization direction of tetragonal BTO was determined from the relative displacement of ions in the optimized structure\citep{pol-bto}.
 The crystal structure of optimized SCOF is shown in Fig.\ \ref{fig1}(a).
 \par Next we modeled three different BTO-SCOF-BTO multilayer heterostructures stacked along the [001] direction by using optimized SCOF and BTO structures. The BTO slab was prepared using a 1$\times$1$\times$4 supercell of the optimized bulk BTO unit cell. The cubic and tetragonal structure of BTO were used in the nonpolar (NP) and polar states, respectively,  where the direction of BTO polarization is towards (P1) and away (P2) from the SCOF, as shown in Figs.\ \ref{fig1}(c),(d). The structurally stable SrO-TiO$_2$ interface \citep{Hotta2007,Shin2017} with optimized interlayer distance was chosen in all three  heterostructures. A vacuum layer of 8 {\AA} was used in  all multilayer's unit cells, with a BaO surface termination.
The atomic positions of all atoms in the BTO layers on both sides of the heterostructure were kept fixed during our simulation to provide the effect of a charged environment due to a polarized electrode, using carefully converged calculations \citep{calcs}.

The optimized geometries of bulk SCOF and BTO-SCOF-BTO heterostructures are shown in Fig.\ \ref{fig1}(a-d). The calculated ground state of SCOF is a tetragonal structure with lattice parameters ${a} = {b} =3.92$ {\AA} and ${c}=13.12$ {\AA} which are 
close to the reported experimental values (${a}={b}=3.83$ {\AA} and ${c}=13.21$ {\AA}) \citep{Tsujimoto2016}. There are two Co atoms in the unit cell of SCOF,  both  prefer to be in the HS configuration with magnetic moment 3 $\mu_B$ in the ground state structure (see below). However, in  BTO-SCOF-BTO heterostructures with different orientation of the BTO polarization, the local environment of the Co atoms has significantly changed which leads to a change in the spin state of the Co atoms. 
The theoretically obtained magnetic moment of the two Co atoms in the bulk SCOF structure and in the heterostructures with NP, P1, and P2 configurations are tabulated in Table \ref{tab:table1}. 

In the bulk phase of SCOF the high spin of a Co atom can be realized from the electronic configuration of the Co$^{3+}$ ion, where one of $t_{2g}$ orbitals is completely filled and other two $t_{2g}$ along with $e_g$ orbitals are partially filled. We computed that the  total magnetic moment on each Co atom is 3.1 $\mu_B$. Nominally, a Co$^{3+}$ ion would have a $d^6$ configuration where the magnetic moment  in HS state would be expected to be 4.0 $\mu_B$. Here however the calculated moment is  reduced due to a strong covalence interaction with the nearest neighbor O and F atoms. A significant charge transfer to the Co atom results from this,  which makes the Co-$d$ occupancy closer to 7 electrons in each considered configuration. 
Our calculation shows that the magnetic moment of the Co atoms in the NP configuration are not altered, both Co1 and Co2 magnetic moments are in the HS state as for  bulk SCOF. However, the total magnetic moment of the unit cell  changes for the different polar P1 and P2  configurations. The maximum change is observed in the P1 alignment for the central Co atom which is farthest from the SCOF-BTO interface, i.e., Co2 in Fig.\ \ref{fig1}(a). 
\begin{table}[b]
\caption{\label{tab:table1}%
\textit{Ab initio} calculated magnetic moments (in $\mu_B$) on the Co atoms in bulk SCOF, and in the NP, P1, and P2 configurations of the heterostructures.
}
\begin{ruledtabular}
\begin{tabular}{lcccc}
\textrm{}&
\textrm{SCOF}&
\textrm{NP}&
\textrm{P1}&
\textrm{P2}\\
\colrule
Co1 & 3.101 & 3.025 & 3.068 & 2.945\\
Co2& 3.101 & 3.059 & -0.051 & 3.010 \\
\end{tabular}
\end{ruledtabular}
\end{table}

In the P1 configuration the magnetic moment on Co2 decreased from 3.059 $\mu_B$ to $-0.051$ $\mu_B$, which demonstrates that the Co2 atom spin has switched from HS to LS ($S=0$). Interestingly, in the P2 configuration, the Co2 atom again switched its spin state from LS 
to HS (3.010 $\mu_B$).
Such a large change in the magnetic behavior due to the change of the orientation of the electric polarization leads to a strong ME coupling in the SCOF. 
We have estimated the surface ME-coupling coefficient ($\alpha_s$) by using 
\begin{equation}
\mu_0\Delta \mathcal{M} = \alpha_s {\mathcal{E}}
\end{equation} 
 where $\mu_0$ is the vacuum permeability, $\Delta \mathcal{M}$ is the change in magnetic flux and $\mathcal{E}$ is the applied electric field. $\Delta \mathcal{M}$ was estimated by considering the magnetic moment difference between the P1 and P2 states per unit surface area, i.e., 2.938 $\mu_B/{a}^2$ (${a}$ is the in-plane lattice constant of the P1 and P2 heterostructures, which is same as optimized tetragonal BTO lattice constant  i.e., ${a}={b}=3.9767$ {\AA}, ${c/a}=1.0383$). Considering the coercive field of BTO ($\mathcal{E}_c$) of 100 kV/cm  \citep{Duan2008} we obtained the ME coefficient $\alpha_s \approx 21.65 \times10^{-10}$ Gcm$^2$/V. 
 
 It is interesting to compare our result  with previously reported $\alpha_s$ values for different systems.
  Duan \textit{et al.}\ showed that in Fe/BTO bilayers the magnetic properties depend, too,  on the orientation of the BTO ferroelectric polarization and estimated $\alpha_s=0.01\times10^{-10}$ Gcm$^2$/V \citep{Duan2006}.
   Similar values were estimated for ultrathin Co and  Ni film on a BTO surface and BFO/CoFe$_2$O$_4$ interfaces \citep{Duan2008,Zavaliche2005}.
  A robust 180$^{\circ}$ switching of the Fe magnetization by an applied electric field has been predicted for an Fe-Au-Fe trilayer on BTO
  \citep{Fechner2012}. 
A significant increase in the ME-coupling constant values were reported for SrTiO$_3$/SrRuO$_3$ ($2\times10^{-10}$ Gcm$^2$/V) \citep{Rondinelli2008}, MnFe$_3$N/BTO ($4\times 10^{-10}$ Gcm$^2$/V), Mn$_2$/TiO$_2$ ($1.2\times10^{-10}$ Gcm$^2$/V) \citep{Lu2015}, Co/PbZr$_{0.2}$Ti$_{0.8}$O$_3$ ($2\times10^{-10}$ Gcm$^2$/V) \citep{Vlasin2016}, Fe$_3$O$_4$/BTO ($0.7\times10^{-10}$ Gcm$^2$/V) \citep{Niranjan2008} and SrRuO$_{3}$/BTO ($2.3\times10^{-10}$ Gcm$^2$/V) \citep{Niranjan2009}. 
   The largest value $\alpha_s$\,=\,$20\times10^{-10}$ Gcm$^2$/V) was reported for an FeO/BTO interface, when a single layer FeO was considered on a BTO surface \citep{Radaelli2014}. 
    Our estimated value of $\alpha_s$ for SOCF/BTO interface exceeds previously reported values, 
    in most cases, by a factor of 10.
 In comparison to the $\alpha_s$ estimation of Ref.\ \citep{Radaelli2014}, it was found there that the direction of the Fe moment in the FeO layer switches its direction, whereas here in the SCOF layer the central Co switches between HS and LS configurations.

\par
\begin{figure}[t!]
\includegraphics[width=0.9\linewidth]{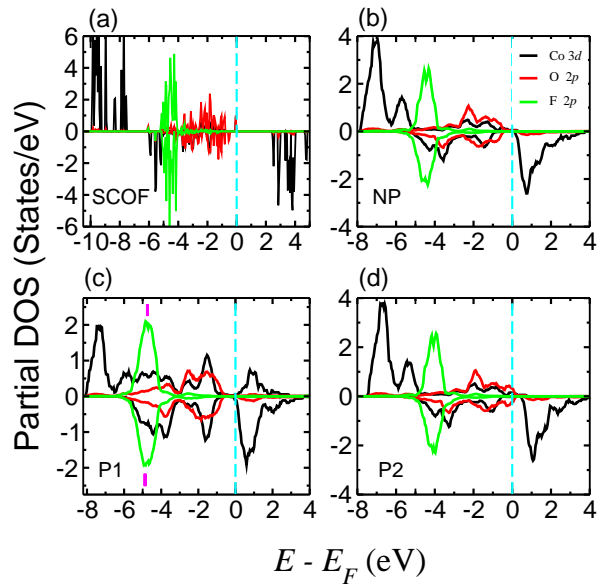}
\vspace{-0.2cm}
\caption{\label{fig2} Atom projected DOS of Co (black), O (red), and F (green lines) in (a) bulk SCOF, (b) NP, (c) P1, and (d) P2 configurations. The solid vertical magenta bars indicate the Co--ligand hybridization which causes the crystal-field splitting in the P1 configuration. Spin-up and spin-down partial DOS are shown by positive and negative values, respectively. 
 The Fermi level ($ E_F$) is set at $E=0$.}
\end{figure}

To understand the spin-state switching mechanisms we carefully studied the electronic structure of the system. Thereto we used the hybrid functional method to obtain an accurate band gap and relative positions of bands. We used a 20\%  Hartree-Fock exchange on top of GGA. The thus-calculated band gap for SCOF in its bulk phase is 2.42 eV. Spin-resolved and atom-projected density of states (DOS) of  SCOF in its bulk phase as well as in BTO-SCOF-BTO heterostructures are shown in Fig.\ \ref{fig2}. 

\begin{figure}[b!]
\vspace{-0.6cm}
\includegraphics[width=0.8\linewidth]{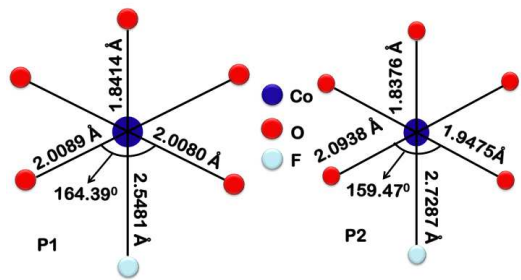}
\vspace{-1.4cm}
\caption{\label{fig3} Calculated bond angle and bond lengths of the Co2 octahedron in the P1 and P2 configurations. The atoms are labelled by the corresponding colors.}
\end{figure}

In all four cases,  both the 
 valance band maxima and the conduction band minima are composed of Co $3d$ and O $2p$ states. 
Note that the bound charge created at the interface due to the polarized BTO may have a significant effect on the local structure of SCOF, specially the anion's position can be expected to be changed due to charge accumulations. The calculated bond lengths and bond angles of the Co2 octahedron in the P1 and P2 configurations are shown in Fig.\ \ref{fig3}.

\begin{figure}[t!]
\includegraphics[width=0.9\linewidth]{PDOS-stack-v2.eps}
\includegraphics[width=6cm]{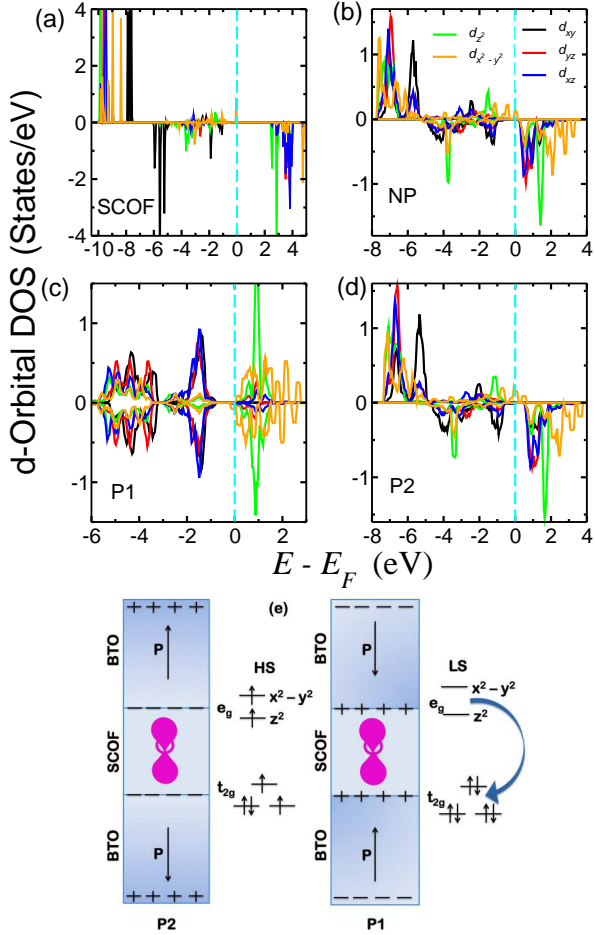}
\caption{\label{fig4} The $3d$-orbital projected DOS of the Co2 atom in (a) bulk SCOF, (b) NP, (c) P1, and (d) P2 configurations. 
The $3d$ orbitals are depicted by the different colors and
spin-up and spin-down DOS are shown as positive and negative values, respectively. (e) Schematic diagram of electron occupancy in the high-spin  and low-spin state configuration of Co2.}
\end{figure}

In the P1 configuration, the polarization is pointing toward the SCOF, therefore there are positive bound charges at the interface.  
We observed that the Co2--F bond length reduced by $\sim$0.2 {\AA} in this configuration compare to the P2. Therefore an increase of covalency in the Co--F bond is expected. Also positive bound charges at the interface affects the electrons in outer orbitals ($d_{xz}$ and $d_{yz}$ states) which leads to a charge redistribution in the $d$-orbitals of the Co2 atom. To understand the effect we further calculated the orbital projected DOS of Co2 $3d$ electrons in bulk SCOF and NP, P1, and P2 heterostructures, shown in Fig.\ \ref{fig4}.
We found that three of the $t_{2g}$ orbitals are fully occupied and the $e_{g}$ orbitals are completely empty in the P1 configuration, whereas they are partially filed in the bulk configuration. This result can be explained by considering the strong crystal-field effect due to induced polarization and internal strain in SCOF. The $2p$ orbital of the F ligand hybridized more with the Co $3d$ orbitals due to the reduction of  Co2--F bond length. Subsequently, the energy level of the Co $d_{z^2}$ orbital is increased (see Fig.\ \ref{fig2}(c)). We also observe that the  O--Co2--O bond angle 
increases, which allows the $d_{x^2-y^2}$ orbital to come closer to the in-plane ligand orbitals and rise its energy level while $d_{xy}$ is away from them. 
 In addition, positive bound charges at the interface lower the energy level of the $d_{ xz}$ and $d_{yz}$ orbitals. Hence, the structural distortion in Co2 octahedron enhances the crystal-field splitting where the $e_{g}$ orbitals of the Co2 atom that were partially occupied in the bulk phase are now  formally unoccupied.  
Empty $d_{x^2-y^2}$ and $d_{z^2}$ orbitals are recognizable in Fig.\ \ref{fig4}(c).
Therefore the electron configuration of the trivalent Co2 atom is approximately $t^{(6)}_{2g}e^{(0)}_g$ and has the LS state.
\par On the other hand in the P2 heterostructure, the polarization is directed away from SCOF. Therefore, the negative bound charge accumulated at the interface that does not effect much on the local environment of the Co2. As a result the change in the crystal field is not significant and the crystal field is very similar to the case of NP configuration. Both $e_{g}$ orbitals of the Co2 atom are now partially field. 
All $d$-orbitals in this case are completely filled with majority spin electrons. The minority spin electrons are present only in the $d_{xy}$ orbital (see Fig.\ \ref{fig4}(d)). Hence, the electron configuration of Co2 in the P2 heterostructure is $\sim t^{(4)}_{2g}e^{(2)}_g$,  and exhibits the HS state. The non-hybridization of the Co2 ion with F ligand orbitals is the witness for the HS state of the Co2 atom in P2 configuration shown in Fig.\ \ref{fig4}(d).
The relative occupation of majority and minority spin electrons in the $3d$-orbitals in the different spin-state configurations are schematically shown in Fig.\ \ref{fig4}(e).
A similar trend is observed in bulk SCOF and the NP configuration. 

To conclude, we have demonstrated the polarization induced spin-state switching and giant magnetoelectric effect in structurally stable perovskite BTO-SCOF-BTO heterostructures. Our calculations show that the magnetic properties of the Co2 atom in a SCOF layer can be tuned by altering the direction of internal polarization of BTO. 
Different bound charges at the interface created due to the polar BTO slab influence significantly the energy levels of Co $3d$ orbitals. As a result, the different polarization states in BTO induce a spin-state transition of the central Co atom from HS to LS and LS to HS state. The spin-state switching of the system due to a polarization change gives rise to a giant ME-coupling constant. Our investigation provides a pathway for exploring new polarization-induced spin switching in oxyfluoride spin-crossover materials and to achieve record magnetoelectric couplings.

We gratefully acknowledge financial support from the Indo-Swedish Research Collaboration, funded through the Swedish Research Council (VR) and the Department of Science and Technology (DST), India. We acknowledge computer time received from the Swedish National Infrastructure for Computing (SNIC).
%

%

\begin{thebibliography}{37}%
\makeatletter
\providecommand \@ifxundefined [1]{%
 \@ifx{#1\undefined}
}%
\providecommand \@ifnum [1]{%
 \ifnum #1\expandafter \@firstoftwo
 \else \expandafter \@secondoftwo
 \fi
}%
\providecommand \@ifx [1]{%
 \ifx #1\expandafter \@firstoftwo
 \else \expandafter \@secondoftwo
 \fi
}%
\providecommand \natexlab [1]{#1}%
\providecommand \enquote  [1]{``#1''}%
\providecommand \bibnamefont  [1]{#1}%
\providecommand \bibfnamefont [1]{#1}%
\providecommand \citenamefont [1]{#1}%
\providecommand \href@noop [0]{\@secondoftwo}%
\providecommand \href [0]{\begingroup \@sanitize@url \@href}%
\providecommand \@href[1]{\@@startlink{#1}\@@href}%
\providecommand \@@href[1]{\endgroup#1\@@endlink}%
\providecommand \@sanitize@url [0]{\catcode `\\12\catcode `\$12\catcode
  `\&12\catcode `\#12\catcode `\^12\catcode `\_12\catcode `\%12\relax}%
\providecommand \@@startlink[1]{}%
\providecommand \@@endlink[0]{}%
\providecommand \url  [0]{\begingroup\@sanitize@url \@url }%
\providecommand \@url [1]{\endgroup\@href {#1}{\urlprefix }}%
\providecommand \urlprefix  [0]{URL }%
\providecommand \Eprint [0]{\href }%
\providecommand \doibase [0]{http://dx.doi.org/}%
\providecommand \selectlanguage [0]{\@gobble}%
\providecommand \bibinfo  [0]{\@secondoftwo}%
\providecommand \bibfield  [0]{\@secondoftwo}%
\providecommand \translation [1]{[#1]}%
\providecommand \BibitemOpen [0]{}%
\providecommand \bibitemStop [0]{}%
\providecommand \bibitemNoStop [0]{.\EOS\space}%
\providecommand \EOS [0]{\spacefactor3000\relax}%
\providecommand \BibitemShut  [1]{\csname bibitem#1\endcsname}%
\let\auto@bib@innerbib\@empty
\bibitem [{\citenamefont {Eerenstein}\ \emph {et~al.}(2006)\citenamefont
  {Eerenstein}, \citenamefont {Mathur},\ and\ \citenamefont
  {Scott}}]{Eerenstein2006}%
  \BibitemOpen
  \bibfield  {author} {\bibinfo {author} {\bibfnamefont {W.}~\bibnamefont
  {Eerenstein}}, \bibinfo {author} {\bibfnamefont {N.~D.}\ \bibnamefont
  {Mathur}}, \ and\ \bibinfo {author} {\bibfnamefont {J.~F.}\ \bibnamefont
  {Scott}},\ }\href {\doibase 10.1038/nature05023} {\bibfield  {journal}
  {\bibinfo  {journal} {Nature}\ }\textbf {\bibinfo {volume} {442}},\ \bibinfo
  {pages} {759} (\bibinfo {year} {2006})}\BibitemShut {NoStop}%
\bibitem [{\citenamefont {Trassin}(2015)}]{Trassin2015}%
  \BibitemOpen
  \bibfield  {author} {\bibinfo {author} {\bibfnamefont {M.}~\bibnamefont
  {Trassin}},\ }\href {\doibase 10.1088/0953-8984/28/3/033001} {\bibfield
  {journal} {\bibinfo  {journal} {J. Phys. Condens. Matter}\ }\textbf {\bibinfo
  {volume} {28}},\ \bibinfo {pages} {33001} (\bibinfo {year}
  {2015})}\BibitemShut {NoStop}%
\bibitem [{\citenamefont {Scott}(2012)}]{Scott2012}%
  \BibitemOpen
  \bibfield  {author} {\bibinfo {author} {\bibfnamefont {J.~F.}\ \bibnamefont
  {Scott}},\ }\href {\doibase 10.1039/c2jm16137k} {\bibfield  {journal}
  {\bibinfo  {journal} {J. Mater. Chem.}\ }\textbf {\bibinfo {volume} {22}},\
  \bibinfo {pages} {4567} (\bibinfo {year} {2012})}\BibitemShut {NoStop}%
\bibitem [{\citenamefont {Hu}\ \emph {et~al.}(2011)\citenamefont {Hu},
  \citenamefont {Li}, \citenamefont {Chen},\ and\ \citenamefont
  {Nan}}]{Hu2011}%
  \BibitemOpen
  \bibfield  {author} {\bibinfo {author} {\bibfnamefont {J.~M.}\ \bibnamefont
  {Hu}}, \bibinfo {author} {\bibfnamefont {Z.}~\bibnamefont {Li}}, \bibinfo
  {author} {\bibfnamefont {L.~Q.}\ \bibnamefont {Chen}}, \ and\ \bibinfo
  {author} {\bibfnamefont {C.~W.}\ \bibnamefont {Nan}},\ }\href {\doibase
  10.1038/ncomms1564} {\bibfield  {journal} {\bibinfo  {journal} {Nat.
  Commun.}\ }\textbf {\bibinfo {volume} {2}},\ \bibinfo {pages} {553} (\bibinfo
  {year} {2011})}\BibitemShut {NoStop}%
\bibitem [{\citenamefont {Hu}\ \emph {et~al.}(2012)\citenamefont {Hu},
  \citenamefont {Li}, \citenamefont {Chen},\ and\ \citenamefont
  {Nan}}]{Hu2012}%
  \BibitemOpen
  \bibfield  {author} {\bibinfo {author} {\bibfnamefont {J.~M.}\ \bibnamefont
  {Hu}}, \bibinfo {author} {\bibfnamefont {Z.}~\bibnamefont {Li}}, \bibinfo
  {author} {\bibfnamefont {L.~Q.}\ \bibnamefont {Chen}}, \ and\ \bibinfo
  {author} {\bibfnamefont {C.~W.}\ \bibnamefont {Nan}},\ }\href {\doibase
  10.1002/adma.201201004} {\bibfield  {journal} {\bibinfo  {journal} {Adv.
  Mater.}\ }\textbf {\bibinfo {volume} {24}},\ \bibinfo {pages} {2869}
  (\bibinfo {year} {2012})}\BibitemShut {NoStop}%
\bibitem [{\citenamefont {Wang}\ \emph {et~al.}(2014)\citenamefont {Wang},
  \citenamefont {Li},\ and\ \citenamefont {Viehland}}]{Wang2014}%
  \BibitemOpen
  \bibfield  {author} {\bibinfo {author} {\bibfnamefont {Y.}~\bibnamefont
  {Wang}}, \bibinfo {author} {\bibfnamefont {J.}~\bibnamefont {Li}}, \ and\
  \bibinfo {author} {\bibfnamefont {D.}~\bibnamefont {Viehland}},\ }\href
  {\doibase 10.1016/j.mattod.2014.05.004} {\bibfield  {journal} {\bibinfo
  {journal} {Materials Today}\ }\textbf {\bibinfo {volume} {17}},\ \bibinfo
  {pages} {269} (\bibinfo {year} {2014})}\BibitemShut {NoStop}%
\bibitem [{\citenamefont {Scott}(2007)}]{Scott2007}%
  \BibitemOpen
  \bibfield  {author} {\bibinfo {author} {\bibfnamefont {J.~F.}\ \bibnamefont
  {Scott}},\ }\href {\doibase 10.1126/science.1129564} {\bibfield  {journal}
  {\bibinfo  {journal} {Science}\ }\textbf {\bibinfo {volume} {315}},\ \bibinfo
  {pages} {954} (\bibinfo {year} {2007})}\BibitemShut {NoStop}%
\bibitem [{\citenamefont {Vaz}\ \emph {et~al.}(2010)\citenamefont {Vaz},
  \citenamefont {Hoffman}, \citenamefont {Ahn},\ and\ \citenamefont
  {Ramesh}}]{Vaz2010}%
  \BibitemOpen
  \bibfield  {author} {\bibinfo {author} {\bibfnamefont {C.~A.~F.}\
  \bibnamefont {Vaz}}, \bibinfo {author} {\bibfnamefont {J.}~\bibnamefont
  {Hoffman}}, \bibinfo {author} {\bibfnamefont {C.~H.}\ \bibnamefont {Ahn}}, \
  and\ \bibinfo {author} {\bibfnamefont {R.}~\bibnamefont {Ramesh}},\ }\href
  {\doibase 10.1002/adma.200904326} {\bibfield  {journal} {\bibinfo  {journal}
  {Adv. Mater.}\ }\textbf {\bibinfo {volume} {22}},\ \bibinfo {pages} {2900}
  (\bibinfo {year} {2010})}\BibitemShut {NoStop}%
\bibitem [{\citenamefont {Scott}(2013)}]{Scott2013}%
  \BibitemOpen
  \bibfield  {author} {\bibinfo {author} {\bibfnamefont {J.~F.}\ \bibnamefont
  {Scott}},\ }\href {\doibase 10.1038/am.2013.58} {\bibfield  {journal}
  {\bibinfo  {journal} {NPG Asia Materials}\ }\textbf {\bibinfo {volume} {5}},\
  \bibinfo {pages} {e72} (\bibinfo {year} {2013})}\BibitemShut {NoStop}%
\bibitem [{\citenamefont {Lu}\ \emph {et~al.}(2015)\citenamefont {Lu},
  \citenamefont {Hu}, \citenamefont {Tian},\ and\ \citenamefont {Wu}}]{Lu2015}%
  \BibitemOpen
  \bibfield  {author} {\bibinfo {author} {\bibfnamefont {C.}~\bibnamefont
  {Lu}}, \bibinfo {author} {\bibfnamefont {W.}~\bibnamefont {Hu}}, \bibinfo
  {author} {\bibfnamefont {Y.}~\bibnamefont {Tian}}, \ and\ \bibinfo {author}
  {\bibfnamefont {T.}~\bibnamefont {Wu}},\ }\href {\doibase 10.1063/1.4921545}
  {\bibfield  {journal} {\bibinfo  {journal} {Appl. Phys. Rev.}\ }\textbf
  {\bibinfo {volume} {2}},\ \bibinfo {pages} {021304} (\bibinfo {year}
  {2015})}\BibitemShut {NoStop}%
\bibitem [{\citenamefont {Ma}\ \emph {et~al.}(2011)\citenamefont {Ma},
  \citenamefont {Hu}, \citenamefont {Li},\ and\ \citenamefont {Nan}}]{Ma2011}%
  \BibitemOpen
  \bibfield  {author} {\bibinfo {author} {\bibfnamefont {J.}~\bibnamefont
  {Ma}}, \bibinfo {author} {\bibfnamefont {J.}~\bibnamefont {Hu}}, \bibinfo
  {author} {\bibfnamefont {Z.}~\bibnamefont {Li}}, \ and\ \bibinfo {author}
  {\bibfnamefont {C.~W.}\ \bibnamefont {Nan}},\ }\href {\doibase
  10.1002/adma.201003636} {\bibfield  {journal} {\bibinfo  {journal} {Adv.
  Mater.}\ }\textbf {\bibinfo {volume} {23}},\ \bibinfo {pages} {1062}
  (\bibinfo {year} {2011})}\BibitemShut {NoStop}%
\bibitem [{\citenamefont {Stroppa}\ \emph {et~al.}(2011)\citenamefont
  {Stroppa}, \citenamefont {Jain}, \citenamefont {Barone}, \citenamefont
  {Marsman}, \citenamefont {Perez-Mato}, \citenamefont {Cheetham},
  \citenamefont {Kroto},\ and\ \citenamefont {Picozzi}}]{Stroppa2011}%
  \BibitemOpen
  \bibfield  {author} {\bibinfo {author} {\bibfnamefont {A.}~\bibnamefont
  {Stroppa}}, \bibinfo {author} {\bibfnamefont {P.}~\bibnamefont {Jain}},
  \bibinfo {author} {\bibfnamefont {P.}~\bibnamefont {Barone}}, \bibinfo
  {author} {\bibfnamefont {M.}~\bibnamefont {Marsman}}, \bibinfo {author}
  {\bibfnamefont {J.~M.}\ \bibnamefont {Perez-Mato}}, \bibinfo {author}
  {\bibfnamefont {A.~K.}\ \bibnamefont {Cheetham}}, \bibinfo {author}
  {\bibfnamefont {H.~W.}\ \bibnamefont {Kroto}}, \ and\ \bibinfo {author}
  {\bibfnamefont {S.}~\bibnamefont {Picozzi}},\ }\href@noop {} {\bibfield
  {journal} {\bibinfo  {journal} {Angew. Chem.}\ }\textbf {\bibinfo {volume}
  {123}},\ \bibinfo {pages} {5969} (\bibinfo {year} {2011})}\BibitemShut
  {NoStop}%
\bibitem [{\citenamefont {{Di Sante}}\ \emph {et~al.}(2013)\citenamefont {{Di
  Sante}}, \citenamefont {Stroppa}, \citenamefont {Jain},\ and\ \citenamefont
  {Picozzi}}]{DiSante2013}%
  \BibitemOpen
  \bibfield  {author} {\bibinfo {author} {\bibfnamefont {D.}~\bibnamefont {{Di
  Sante}}}, \bibinfo {author} {\bibfnamefont {A.}~\bibnamefont {Stroppa}},
  \bibinfo {author} {\bibfnamefont {P.}~\bibnamefont {Jain}}, \ and\ \bibinfo
  {author} {\bibfnamefont {S.}~\bibnamefont {Picozzi}},\ }\href@noop {}
  {\bibfield  {journal} {\bibinfo  {journal} {J. Am. Chem. Soc.}\ }\textbf
  {\bibinfo {volume} {135}},\ \bibinfo {pages} {18126} (\bibinfo {year}
  {2013})}\BibitemShut {NoStop}%
\bibitem [{\citenamefont {Stroppa}\ \emph {et~al.}(2013)\citenamefont
  {Stroppa}, \citenamefont {Barone}, \citenamefont {Jain}, \citenamefont
  {Perez-Mato},\ and\ \citenamefont {Picozzi}}]{Stroppa2013}%
  \BibitemOpen
  \bibfield  {author} {\bibinfo {author} {\bibfnamefont {A.}~\bibnamefont
  {Stroppa}}, \bibinfo {author} {\bibfnamefont {P.}~\bibnamefont {Barone}},
  \bibinfo {author} {\bibfnamefont {P.}~\bibnamefont {Jain}}, \bibinfo {author}
  {\bibfnamefont {J.~M.}\ \bibnamefont {Perez-Mato}}, \ and\ \bibinfo {author}
  {\bibfnamefont {S.}~\bibnamefont {Picozzi}},\ }\href@noop {} {\bibfield
  {journal} {\bibinfo  {journal} {Adv. Mater.}\ }\textbf {\bibinfo {volume}
  {25}},\ \bibinfo {pages} {2284} (\bibinfo {year} {2013})}\BibitemShut
  {NoStop}%
 \bibitem [{\citenamefont {Fechner}\ \emph {et~al.}(2012)\citenamefont {Fechner},
  \citenamefont {Zahn}, \citenamefont {Ostanin}, \citenamefont {Bibes},
  \ and\ \citenamefont
  {Mertig}}]{Fechner2012}%
  \BibitemOpen
  \bibfield  {author} {\bibinfo {author} {\bibfnamefont {C.}\ \bibnamefont
  {Fechner}}, \bibinfo {author} {\bibfnamefont {P.}\ \bibnamefont {Zahn}},
  \bibinfo {author} {\bibfnamefont {S.}\ \bibnamefont {Ostanin}},
  \bibinfo {author} {\bibfnamefont {M.}~\bibnamefont {Bibes}} \ and\ \bibinfo {author} {\bibfnamefont
  {I.}\ \bibnamefont {Mertig}},\ }\href {\doibase
  10.1103/PhysRevLett.108.197206} {\bibfield  {journal} {\bibinfo  {journal}
  {Phys. Rev. Lett.}\ }\textbf {\bibinfo {volume} {108}},\ \bibinfo {pages}
  {197206} (\bibinfo {year} {2012})}\BibitemShut {NoStop}%
\bibitem [{\citenamefont {Moritomo}\ \emph {et~al.}(2000)\citenamefont
  {Moritomo}, \citenamefont {Akimoto}, \citenamefont {Takeo}, \citenamefont
  {Machida}, \citenamefont {Nishibori}, \citenamefont {Takata}, \citenamefont
  {Sakata}, \citenamefont {Ohoyama},\ and\ \citenamefont
  {Nakamura}}]{Moritomo2000}%
  \BibitemOpen
  \bibfield  {author} {\bibinfo {author} {\bibfnamefont {Y.}~\bibnamefont
  {Moritomo}}, \bibinfo {author} {\bibfnamefont {T.}~\bibnamefont {Akimoto}},
  \bibinfo {author} {\bibfnamefont {M.}~\bibnamefont {Takeo}}, \bibinfo
  {author} {\bibfnamefont {A.}~\bibnamefont {Machida}}, \bibinfo {author}
  {\bibfnamefont {E.}~\bibnamefont {Nishibori}}, \bibinfo {author}
  {\bibfnamefont {M.}~\bibnamefont {Takata}}, \bibinfo {author} {\bibfnamefont
  {M.}~\bibnamefont {Sakata}}, \bibinfo {author} {\bibfnamefont
  {K.}~\bibnamefont {Ohoyama}}, \ and\ \bibinfo {author} {\bibfnamefont
  {A.}~\bibnamefont {Nakamura}},\ }\href {\doibase 10.1103/PhysRevB.61.R13325}
  {\bibfield  {journal} {\bibinfo  {journal} {Phys. Rev. B}\ }\textbf {\bibinfo
  {volume} {61}},\ \bibinfo {pages} {R13325} (\bibinfo {year}
  {2000})}\BibitemShut {NoStop}%
\bibitem [{\citenamefont {Briceno}\ \emph {et~al.}(2018)\citenamefont
  {Briceno}, \citenamefont {Chang},\ and\ \citenamefont {Sun}}]{Briceno2018}%
  \BibitemOpen
  \bibfield  {author} {\bibinfo {author} {\bibfnamefont {G.}~\bibnamefont
  {Briceno}}, \bibinfo {author} {\bibfnamefont {H.}~\bibnamefont {Chang}}, \
  and\ \bibinfo {author} {\bibfnamefont {X.}~\bibnamefont {Sun}},\ }\href@noop
  {} {\ \textbf {\bibinfo {volume} {270}},\ \bibinfo {pages} {273} (\bibinfo
  {year} {2018})}\BibitemShut {NoStop}%
\bibitem [{\citenamefont {Tsujimoto}\ \emph {et~al.}(2012)\citenamefont
  {Tsujimoto}, \citenamefont {Sathish}, \citenamefont {Hong}, \citenamefont
  {Oka}, \citenamefont {Azuma}, \citenamefont {Guo}, \citenamefont
  {Matsushita}, \citenamefont {Yamaura},\ and\ \citenamefont
  {Takayama-Muromachi}}]{Oxyfluorides2012}%
  \BibitemOpen
  \bibfield  {author} {\bibinfo {author} {\bibfnamefont {Y.}~\bibnamefont
  {Tsujimoto}}, \bibinfo {author} {\bibfnamefont {C.~I.}\ \bibnamefont
  {Sathish}}, \bibinfo {author} {\bibfnamefont {K.-p.}\ \bibnamefont {Hong}},
  \bibinfo {author} {\bibfnamefont {K.}~\bibnamefont {Oka}}, \bibinfo {author}
  {\bibfnamefont {M.}~\bibnamefont {Azuma}}, \bibinfo {author} {\bibfnamefont
  {Y.}~\bibnamefont {Guo}}, \bibinfo {author} {\bibfnamefont {Y.}~\bibnamefont
  {Matsushita}}, \bibinfo {author} {\bibfnamefont {K.}~\bibnamefont {Yamaura}},
  \ and\ \bibinfo {author} {\bibfnamefont {E.}~\bibnamefont
  {Takayama-Muromachi}},\ }\href@noop {} {\bibfield  {journal} {\bibinfo
  {journal} {Inorg. Chem.}\ }\textbf {\bibinfo {volume} {51}},\ \bibinfo
  {pages} {4802} (\bibinfo {year} {2012})}\BibitemShut {NoStop}%
\bibitem [{\citenamefont {Tsujimoto}\ \emph {et~al.}(2016)\citenamefont
  {Tsujimoto}, \citenamefont {Nakano}, \citenamefont {Ishimatsu}, \citenamefont
  {Mizumaki}, \citenamefont {Kawamura}, \citenamefont {Kawakami}, \citenamefont
  {Matsushita},\ and\ \citenamefont {Yamaura}}]{Tsujimoto2016}%
  \BibitemOpen
  \bibfield  {author} {\bibinfo {author} {\bibfnamefont {Y.}~\bibnamefont
  {Tsujimoto}}, \bibinfo {author} {\bibfnamefont {S.}~\bibnamefont {Nakano}},
  \bibinfo {author} {\bibfnamefont {N.}~\bibnamefont {Ishimatsu}}, \bibinfo
  {author} {\bibfnamefont {M.}~\bibnamefont {Mizumaki}}, \bibinfo {author}
  {\bibfnamefont {N.}~\bibnamefont {Kawamura}}, \bibinfo {author}
  {\bibfnamefont {T.}~\bibnamefont {Kawakami}}, \bibinfo {author}
  {\bibfnamefont {Y.}~\bibnamefont {Matsushita}}, \ and\ \bibinfo {author}
  {\bibfnamefont {K.}~\bibnamefont {Yamaura}},\ }\href {\doibase
  10.1038/srep36253} {\bibfield  {journal} {\bibinfo  {journal} {Sci. Rep.}\
  }\textbf {\bibinfo {volume} {6}},\ \bibinfo {pages} {36253} (\bibinfo {year}
  {2016})}\BibitemShut {NoStop}%
\bibitem [{\citenamefont {Ou}\ \emph {et~al.}(2016)\citenamefont {Ou},
  \citenamefont {Fan}, \citenamefont {Li}, \citenamefont {Wang},\ and\
  \citenamefont {Wu}}]{Ou2016}%
  \BibitemOpen
  \bibfield  {author} {\bibinfo {author} {\bibfnamefont {X.}~\bibnamefont
  {Ou}}, \bibinfo {author} {\bibfnamefont {F.}~\bibnamefont {Fan}}, \bibinfo
  {author} {\bibfnamefont {Z.}~\bibnamefont {Li}}, \bibinfo {author}
  {\bibfnamefont {H.}~\bibnamefont {Wang}}, \ and\ \bibinfo {author}
  {\bibfnamefont {H.}~\bibnamefont {Wu}},\ }\href {\doibase 10.1063/1.4943104}
  {\bibfield  {journal} {\bibinfo  {journal} {Appl. Phys. Lett.}\ }\textbf
  {\bibinfo {volume} {108}},\ \bibinfo {pages} {092402} (\bibinfo {year}
  {2016})}\BibitemShut {NoStop}%
\bibitem [{\citenamefont {Kresse}\ and\ \citenamefont
  {Furthm{\"{u}}ller}(1996)}]{Kresse1996}%
  \BibitemOpen
  \bibfield  {author} {\bibinfo {author} {\bibfnamefont {G.}~\bibnamefont
  {Kresse}}\ and\ \bibinfo {author} {\bibfnamefont {J.}~\bibnamefont
  {Furthm{\"{u}}ller}},\ }\href
  {http://link.aps.org/doi/10.1103/PhysRevB.54.11169} {\bibfield  {journal}
  {\bibinfo  {journal} {Phys. Rev. B}\ }\textbf {\bibinfo {volume} {54}},\
  \bibinfo {pages} {11169} (\bibinfo {year} {1996})}\BibitemShut {NoStop}%
\bibitem [{\citenamefont {Perdew}\ \emph {et~al.}(1996)\citenamefont {Perdew},
  \citenamefont {Burke},\ and\ \citenamefont {Ernzerhof}}]{Perdew1996}%
  \BibitemOpen
  \bibfield  {author} {\bibinfo {author} {\bibfnamefont {J.~P.}\ \bibnamefont
  {Perdew}}, \bibinfo {author} {\bibfnamefont {K.}~\bibnamefont {Burke}}, \
  and\ \bibinfo {author} {\bibfnamefont {M.}~\bibnamefont {Ernzerhof}},\ }\href
  {\doibase 10.1103/PhysRevLett.77.3865} {\bibfield  {journal} {\bibinfo
  {journal} {Phys. Rev. Lett.}\ }\textbf {\bibinfo {volume} {77}},\ \bibinfo
  {pages} {3865} (\bibinfo {year} {1996})}\BibitemShut {NoStop}%
\bibitem [{\citenamefont {Dudarev}\ \emph {et~al.}(1998)\citenamefont
  {Dudarev}, \citenamefont {Botton}, \citenamefont {Savrasov}, \citenamefont
  {Humphreys},\ and\ \citenamefont {Sutton}}]{Dudarev1998}%
  \BibitemOpen
  \bibfield  {author} {\bibinfo {author} {\bibfnamefont {S.~L.}\ \bibnamefont
  {Dudarev}}, \bibinfo {author} {\bibfnamefont {G.~A.}\ \bibnamefont {Botton}},
  \bibinfo {author} {\bibfnamefont {S.~Y.}\ \bibnamefont {Savrasov}}, \bibinfo
  {author} {\bibfnamefont {C.~J.}\ \bibnamefont {Humphreys}}, \ and\ \bibinfo
  {author} {\bibfnamefont {A.~P.}\ \bibnamefont {Sutton}},\ }\href {\doibase
  10.1103/PhysRevB.57.1505} {\bibfield  {journal} {\bibinfo  {journal} {Phys.
  Rev. B}\ }\textbf {\bibinfo {volume} {57}},\ \bibinfo {pages} {1505}
  (\bibinfo {year} {1998})}\BibitemShut {NoStop}%
\bibitem [{\citenamefont {Sarkar}\ \emph {et~al.}(2011)\citenamefont {Sarkar},
  \citenamefont {Tarafder}, \citenamefont {Oppeneer},\ and\ \citenamefont
  {Saha-Dasgupta}}]{Sarkar2011}%
  \BibitemOpen
  \bibfield  {author} {\bibinfo {author} {\bibfnamefont {S.}~\bibnamefont
  {Sarkar}}, \bibinfo {author} {\bibfnamefont {K.}~\bibnamefont {Tarafder}},
  \bibinfo {author} {\bibfnamefont {P.~M.}\ \bibnamefont {Oppeneer}}, \ and\
  \bibinfo {author} {\bibfnamefont {T.}~\bibnamefont {Saha-Dasgupta}},\ }\href
  {\doibase 10.1039/C1JM11679G} {\bibfield  {journal} {\bibinfo  {journal} {J.
  Mater. Chem.}\ }\textbf {\bibinfo {volume} {21}},\ \bibinfo {pages} {13832}
  (\bibinfo {year} {2011})}\BibitemShut {NoStop}%
\bibitem [{\citenamefont {Tarafder}\ \emph {et~al.}(2012)\citenamefont
  {Tarafder}, \citenamefont {Kanungo}, \citenamefont {Oppeneer},\ and\
  \citenamefont {Saha-Dasgupta}}]{Tarafder2012}%
  \BibitemOpen
  \bibfield  {author} {\bibinfo {author} {\bibfnamefont {K.}~\bibnamefont
  {Tarafder}}, \bibinfo {author} {\bibfnamefont {S.}~\bibnamefont {Kanungo}},
  \bibinfo {author} {\bibfnamefont {P.~M.}\ \bibnamefont {Oppeneer}}, \ and\
  \bibinfo {author} {\bibfnamefont {T.}~\bibnamefont {Saha-Dasgupta}},\ }\href
  {\doibase 10.1103/PhysRevLett.109.077203} {\bibfield  {journal} {\bibinfo
  {journal} {Phys. Rev. Lett.}\ }\textbf {\bibinfo {volume} {109}},\ \bibinfo
  {pages} {077203} (\bibinfo {year} {2012})}\BibitemShut {NoStop}%
\bibitem [{\citenamefont {Maldonado}\ \emph {et~al.}(2013)\citenamefont
  {Maldonado}, \citenamefont {Kanungo}, \citenamefont {Saha-Dasgupta},\ and\
  \citenamefont {Oppeneer}}]{Maldonado2013}%
  \BibitemOpen
  \bibfield  {author} {\bibinfo {author} {\bibfnamefont {P.}~\bibnamefont
  {Maldonado}}, \bibinfo {author} {\bibfnamefont {S.}~\bibnamefont {Kanungo}},
  \bibinfo {author} {\bibfnamefont {T.}~\bibnamefont {Saha-Dasgupta}}, \ and\
  \bibinfo {author} {\bibfnamefont {P.~M.}\ \bibnamefont {Oppeneer}},\ }\href
  {\doibase 10.1103/PhysRevB.88.020408} {\bibfield  {journal} {\bibinfo
  {journal} {Phys. Rev. B}\ }\textbf {\bibinfo {volume} {88}},\ \bibinfo
  {pages} {020408(R)} (\bibinfo {year} {2013})}\BibitemShut {NoStop}%
\bibitem [{cal()}]{calcs}%
  \BibitemOpen
  \href@noop {} {}\bibinfo {note} {The electron-ionic-core interaction on the
  valence electrons was represented by projector augmented wave potentials
  (PAWs) \cite{Blochl1994}. In all simulations for the heterostructures a
  plane-wave cutoff energy of 400 eV was used. The reciprocal space was sampled
  according to the Monkhorst-Pack scheme and the convergence criterion was set
  to 10$^{-5}$ eV for the self-consistent electronic energy minimization. Full
  structural optimization of all bulk structures was carried out using a
  4$\times$4$\times$3 $k$-point mesh until all residual interatomic forces were
  reduced to less than 0.01 eV/{\AA}. Calculations of the polarized electrode's
  effect on the SCOF layer were done by minimizing the total energy of the
  whole system using a 11$\times$11$\times$1 $k$-point mesh until all residual
  interatomic forces reduced to less than 0.01 eV/{\AA}.}\BibitemShut {Stop}%
\bibitem [{\citenamefont {Wang}\ \emph {et~al.}(2010)\citenamefont {Wang},
  \citenamefont {Meng}, \citenamefont {Ma}, \citenamefont {Xu},\ and\
  \citenamefont {Chen}}]{Wang2010}%
  \BibitemOpen
  \bibfield  {author} {\bibinfo {author} {\bibfnamefont {J.~J.}\ \bibnamefont
  {Wang}}, \bibinfo {author} {\bibfnamefont {F.~Y.}\ \bibnamefont {Meng}},
  \bibinfo {author} {\bibfnamefont {X.~Q.}\ \bibnamefont {Ma}}, \bibinfo
  {author} {\bibfnamefont {M.~X.}\ \bibnamefont {Xu}}, \ and\ \bibinfo {author}
  {\bibfnamefont {L.~Q.}\ \bibnamefont {Chen}},\ }\href {\doibase
  10.1063/1.3462441} {\bibfield  {journal} {\bibinfo  {journal} {J. Appl.
  Phys.}\ }\textbf {\bibinfo {volume} {108}},\ \bibinfo {pages} {034107}
  (\bibinfo {year} {2010})}\BibitemShut {NoStop}%
\bibitem [{\citenamefont {Megaw}(1962)}]{Megaw1962}%
  \BibitemOpen
  \bibfield  {author} {\bibinfo {author} {\bibfnamefont {H.~D.}\ \bibnamefont
  {Megaw}},\ }\href {\doibase 10.1107/S0365110X62002571} {\bibfield  {journal}
  {\bibinfo  {journal} {Acta Crystallographica}\ }\textbf {\bibinfo {volume}
  {15}},\ \bibinfo {pages} {972} (\bibinfo {year} {1962})}\BibitemShut
  {NoStop}%
  \bibitem [{cal()}]{pol-bto}%
  \BibitemOpen
  \href@noop {} {}\bibinfo {note} {The calculated value of spontaneous polarization in tetragonal BaTiO$_3$ is 0.4125 C/m$^2$ .} \BibitemShut {Stop}%
\bibitem [{\citenamefont {Hotta}\ \emph {et~al.}(2007)\citenamefont {Hotta},
  \citenamefont {Suasaki},\ and\ \citenamefont
  {Hwang}}]{Hotta2007}%
  \BibitemOpen
  \bibfield  {author} {\bibinfo {author} {\bibfnamefont {Y.}\ \bibnamefont
  {Hotta}}, \bibinfo {author} {\bibfnamefont {T.}\ \bibnamefont {Suasaki}},
   \ and\ \bibinfo {author} {\bibfnamefont
  {H.~Y.}\ \bibnamefont {Hwang}},\ }\href {\doibase
  10.1103/PhysRevLett.99.236805} {\bibfield  {journal} {\bibinfo  {journal}
  {Phys. Rev. Lett.}\ }\textbf {\bibinfo {volume} {99}},\ \bibinfo {pages}
  {236805} (\bibinfo {year} {2007})}\BibitemShut {NoStop}%
  \bibitem [{\citenamefont {Shin}\ \emph {et~al.}(2017)\citenamefont {Shin},
  \citenamefont {Kim}, \citenamefont {Kang}, \citenamefont {Nahm},
  \citenamefont {Murugavel}, \citenamefont {Kim}, \citenamefont {Cho},
  \citenamefont {Wang}, \citenamefont {Yang},\ and\ \citenamefont
  {Yoon}}]{Shin2017}%
 \BibitemOpen
  \bibfield  {author} {\bibinfo {author}
  {\bibfnamefont {Y.~J.}\ \bibnamefont  {Shin}}, \bibinfo {author} 
  {\bibfnamefont {Y.}\ \bibnamefont {Kim}}, \bibinfo  {author} 
  {\bibfnamefont {S.-J.}\ \bibnamefont {Kang}}, \bibinfo {author}
  {\bibfnamefont {H.-H}\ \bibnamefont {Nahm}}, \bibinfo {author} 
  {\bibfnamefont {P.}\ \bibnamefont {Murugavel}}, \bibinfo {author} 
  {\bibfnamefont {J.~R.}\ \bibnamefont {Kim}}, \bibinfo {author} 
  {\bibfnamefont {M.~R.}\ \bibnamefont   {Cho}}, \bibinfo {author} 
  {\bibfnamefont {L.}\ \bibnamefont {Wang}}, \bibinfo  {author} 
  {\bibfnamefont {S.~M.}\ \bibnamefont {Yang}}, \ and\ \bibinfo  {author} 
  {\bibfnamefont {J.~G.}\ \bibnamefont {Yoon}},\ }\href {\doibase
  10.1002/adma.201602795} {\bibfield  {journal} {\bibinfo  {journal} {Adv.
  Mater.}\ }\textbf {\bibinfo {volume} {29}},\ \bibinfo {pages} {1602795}
  (\bibinfo {year} {2017})}\BibitemShut {NoStop}%
\bibitem [{\citenamefont {Duan}\ \emph {et~al.}(2008)\citenamefont {Duan},
  \citenamefont {Velev}, \citenamefont {Sabirianov}, \citenamefont {Zhu},
  \citenamefont {Chu}, \citenamefont {Jaswal},\ and\ \citenamefont
  {Tsymbal}}]{Duan2008}%
  \BibitemOpen
  \bibfield  {author} {\bibinfo {author} {\bibfnamefont {C.~G.}\ \bibnamefont
  {Duan}}, \bibinfo {author} {\bibfnamefont {J.~P.}\ \bibnamefont {Velev}},
  \bibinfo {author} {\bibfnamefont {R.~F.}\ \bibnamefont {Sabirianov}},
  \bibinfo {author} {\bibfnamefont {Z.}~\bibnamefont {Zhu}}, \bibinfo {author}
  {\bibfnamefont {J.}~\bibnamefont {Chu}}, \bibinfo {author} {\bibfnamefont
  {S.~S.}\ \bibnamefont {Jaswal}}, \ and\ \bibinfo {author} {\bibfnamefont
  {E.~Y.}\ \bibnamefont {Tsymbal}},\ }\href {\doibase
  10.1103/PhysRevLett.101.137201} {\bibfield  {journal} {\bibinfo  {journal}
  {Phys. Rev. Lett.}\ }\textbf {\bibinfo {volume} {101}},\ \bibinfo {pages}
  {137201} (\bibinfo {year} {2008})}\BibitemShut {NoStop}%
\bibitem [{\citenamefont {Duan}\ \emph {et~al.}(2006)\citenamefont {Duan},
  \citenamefont {Jaswal},\ and\ \citenamefont {Tsymbal}}]{Duan2006}%
  \BibitemOpen
  \bibfield  {author} {\bibinfo {author} {\bibfnamefont {C.-G.}\ \bibnamefont
  {Duan}}, \bibinfo {author} {\bibfnamefont {S.~S.}\ \bibnamefont {Jaswal}}, \
  and\ \bibinfo {author} {\bibfnamefont {E.~Y.}\ \bibnamefont {Tsymbal}},\
  }\href {\doibase 10.1103/PhysRevLett.97.047201} {\bibfield  {journal}
  {\bibinfo  {journal} {Phys. Rev. Lett.}\ }\textbf {\bibinfo {volume} {97}},\
  \bibinfo {pages} {047201} (\bibinfo {year} {2006})}\BibitemShut {NoStop}%
\bibitem [{\citenamefont {Zavaliche}\ \emph {et~al.}(2005)\citenamefont
  {Zavaliche}, \citenamefont {Zheng}, \citenamefont {Mohaddes-Ardabili},
  \citenamefont {Yang}, \citenamefont {Zhan}, \citenamefont {Shafer},
  \citenamefont {Reilly}, \citenamefont {Chopdekar}, \citenamefont {Jia},
  \citenamefont {Wright}, \citenamefont {Schlom}, \citenamefont {Suzuki},\ and\
  \citenamefont {Ramesh}}]{Zavaliche2005}%
  \BibitemOpen
  \bibfield  {author} {\bibinfo {author} {\bibfnamefont {F.}~\bibnamefont
  {Zavaliche}}, \bibinfo {author} {\bibfnamefont {H.}~\bibnamefont {Zheng}},
  \bibinfo {author} {\bibfnamefont {L.}~\bibnamefont {Mohaddes-Ardabili}},
  \bibinfo {author} {\bibfnamefont {S.~Y.}\ \bibnamefont {Yang}}, \bibinfo
  {author} {\bibfnamefont {Q.}~\bibnamefont {Zhan}}, \bibinfo {author}
  {\bibfnamefont {P.}~\bibnamefont {Shafer}}, \bibinfo {author} {\bibfnamefont
  {E.}~\bibnamefont {Reilly}}, \bibinfo {author} {\bibfnamefont
  {R.}~\bibnamefont {Chopdekar}}, \bibinfo {author} {\bibfnamefont
  {Y.}~\bibnamefont {Jia}}, \bibinfo {author} {\bibfnamefont {P.}~\bibnamefont
  {Wright}}, \bibinfo {author} {\bibfnamefont {D.~G.}\ \bibnamefont {Schlom}},
  \bibinfo {author} {\bibfnamefont {Y.}~\bibnamefont {Suzuki}}, \ and\ \bibinfo
  {author} {\bibfnamefont {R.}~\bibnamefont {Ramesh}},\ }\href {\doibase
  10.1021/nl051406i} {\bibfield  {journal} {\bibinfo  {journal} {Nano Lett.}\
  }\textbf {\bibinfo {volume} {5}},\ \bibinfo {pages} {1793} (\bibinfo {year}
  {2005})}\BibitemShut {NoStop}%
\bibitem [{\citenamefont {Rondinelli}\ \emph {et~al.}(2008)\citenamefont
  {Rondinelli}, \citenamefont {Stengel},\ and\ \citenamefont
  {Spaldin}}]{Rondinelli2008}%
  \BibitemOpen
  \bibfield  {author} {\bibinfo {author} {\bibfnamefont {J.~M.}\ \bibnamefont
  {Rondinelli}}, \bibinfo {author} {\bibfnamefont {M.}~\bibnamefont {Stengel}},
  \ and\ \bibinfo {author} {\bibfnamefont {N.~A.}\ \bibnamefont {Spaldin}},\
  }\href {\doibase 10.1038/nnano.2007.412} {\bibfield  {journal} {\bibinfo
  {journal} {Nat. Nanotech.}\ }\textbf {\bibinfo {volume} {3}},\ \bibinfo
  {pages} {46} (\bibinfo {year} {2008})}\BibitemShut {NoStop}%
\bibitem [{\citenamefont {Vla{\v{s}}{\'{i}}n}\ \emph
  {et~al.}(2016)\citenamefont {Vla{\v{s}}{\'{i}}n}, \citenamefont {Jarrier},
  \citenamefont {Arras}, \citenamefont {Calmels}, \citenamefont
  {Warot-Fonrose}, \citenamefont {Marcelot}, \citenamefont {Jamet},
  \citenamefont {Ohresser}, \citenamefont {Scheurer}, \citenamefont {Hertel},
  \citenamefont {Herranz},\ and\ \citenamefont {Cherifi-Hertel}}]{Vlasin2016}%
  \BibitemOpen
  \bibfield  {author} {\bibinfo {author} {\bibfnamefont {O.}~\bibnamefont
  {Vla{\v{s}}{\'{i}}n}}, \bibinfo {author} {\bibfnamefont {R.}~\bibnamefont
  {Jarrier}}, \bibinfo {author} {\bibfnamefont {R.}~\bibnamefont {Arras}},
  \bibinfo {author} {\bibfnamefont {L.}~\bibnamefont {Calmels}}, \bibinfo
  {author} {\bibfnamefont {B.}~\bibnamefont {Warot-Fonrose}}, \bibinfo {author}
  {\bibfnamefont {C.}~\bibnamefont {Marcelot}}, \bibinfo {author}
  {\bibfnamefont {M.}~\bibnamefont {Jamet}}, \bibinfo {author} {\bibfnamefont
  {P.}~\bibnamefont {Ohresser}}, \bibinfo {author} {\bibfnamefont
  {F.}~\bibnamefont {Scheurer}}, \bibinfo {author} {\bibfnamefont
  {R.}~\bibnamefont {Hertel}}, \bibinfo {author} {\bibfnamefont
  {G.}~\bibnamefont {Herranz}}, \ and\ \bibinfo {author} {\bibfnamefont
  {S.}~\bibnamefont {Cherifi-Hertel}},\ }\href {\doibase
  10.1021/acsami.5b12777} {\bibfield  {journal} {\bibinfo  {journal} {ACS Appl.
  Mater. Interfaces}\ }\textbf {\bibinfo {volume} {8}},\ \bibinfo {pages}
  {7553} (\bibinfo {year} {2016})}\BibitemShut {NoStop}%
\bibitem [{\citenamefont {Niranjan}\ \emph {et~al.}(2008)\citenamefont
  {Niranjan}, \citenamefont {Velev}, \citenamefont {Duan}, \citenamefont
  {Jaswal},\ and\ \citenamefont {Tsymbal}}]{Niranjan2008}%
  \BibitemOpen
  \bibfield  {author} {\bibinfo {author} {\bibfnamefont {M.~K.}\ \bibnamefont
  {Niranjan}}, \bibinfo {author} {\bibfnamefont {J.~P.}\ \bibnamefont {Velev}},
  \bibinfo {author} {\bibfnamefont {C.~G.}\ \bibnamefont {Duan}}, \bibinfo
  {author} {\bibfnamefont {S.~S.}\ \bibnamefont {Jaswal}}, \ and\ \bibinfo
  {author} {\bibfnamefont {E.~Y.}\ \bibnamefont {Tsymbal}},\ }\href {\doibase
  10.1103/PhysRevB.78.104405} {\bibfield  {journal} {\bibinfo  {journal} {Phys.
  Rev. B}\ }\textbf {\bibinfo {volume} {78}},\ \bibinfo {pages} {104405}
  (\bibinfo {year} {2008})}\BibitemShut {NoStop}%
\bibitem [{\citenamefont {Niranjan}\ \emph {et~al.}(2009)\citenamefont
  {Niranjan}, \citenamefont {Burton}, \citenamefont {Velev}, \citenamefont
  {Jaswal},\ and\ \citenamefont {Tsymbal}}]{Niranjan2009}%
  \BibitemOpen
  \bibfield  {author} {\bibinfo {author} {\bibfnamefont {M.~K.}\ \bibnamefont
  {Niranjan}}, \bibinfo {author} {\bibfnamefont {J.~D.}\ \bibnamefont
  {Burton}}, \bibinfo {author} {\bibfnamefont {J.~P.}\ \bibnamefont {Velev}},
  \bibinfo {author} {\bibfnamefont {S.~S.}\ \bibnamefont {Jaswal}}, \ and\
  \bibinfo {author} {\bibfnamefont {E.~Y.}\ \bibnamefont {Tsymbal}},\ }\href
  {\doibase 10.1063/1.3193679} {\bibfield  {journal} {\bibinfo  {journal}
  {Appl. Phys. Lett.}\ }\textbf {\bibinfo {volume} {95}},\ \bibinfo {pages}
  {052051} (\bibinfo {year} {2009})}\BibitemShut {NoStop}%
\bibitem [{\citenamefont {Radaelli}\ \emph {et~al.}(2014)\citenamefont
  {Radaelli}, \citenamefont {Petti}, \citenamefont {Plekhanov}, \citenamefont
  {Fina}, \citenamefont {Torelli}, \citenamefont {Salles}, \citenamefont
  {Cantoni}, \citenamefont {Rinaldi}, \citenamefont {Guti{\'{e}}rrez},
  \citenamefont {Panaccione}, \citenamefont {Varela}, \citenamefont {Picozzi},
  \citenamefont {Fontcuberta},\ and\ \citenamefont {Bertacco}}]{Radaelli2014}%
  \BibitemOpen
  \bibfield  {author} {\bibinfo {author} {\bibfnamefont {G.}~\bibnamefont
  {Radaelli}}, \bibinfo {author} {\bibfnamefont {D.}~\bibnamefont {Petti}},
  \bibinfo {author} {\bibfnamefont {E.}~\bibnamefont {Plekhanov}}, \bibinfo
  {author} {\bibfnamefont {I.}~\bibnamefont {Fina}}, \bibinfo {author}
  {\bibfnamefont {P.}~\bibnamefont {Torelli}}, \bibinfo {author} {\bibfnamefont
  {B.~R.}\ \bibnamefont {Salles}}, \bibinfo {author} {\bibfnamefont
  {M.}~\bibnamefont {Cantoni}}, \bibinfo {author} {\bibfnamefont
  {C.}~\bibnamefont {Rinaldi}}, \bibinfo {author} {\bibfnamefont
  {D.}~\bibnamefont {Guti{\'{e}}rrez}}, \bibinfo {author} {\bibfnamefont
  {G.}~\bibnamefont {Panaccione}}, \bibinfo {author} {\bibfnamefont
  {M.}~\bibnamefont {Varela}}, \bibinfo {author} {\bibfnamefont
  {S.}~\bibnamefont {Picozzi}}, \bibinfo {author} {\bibfnamefont
  {J.}~\bibnamefont {Fontcuberta}}, \ and\ \bibinfo {author} {\bibfnamefont
  {R.}~\bibnamefont {Bertacco}},\ }\href {\doibase 10.1038/ncomms4404}
  {\bibfield  {journal} {\bibinfo  {journal} {Nat. Commun.}\ }\textbf {\bibinfo
  {volume} {5}},\ \bibinfo {pages} {3404} (\bibinfo {year} {2014})}\BibitemShut
  {NoStop}%
\bibitem [{\citenamefont {Bl{\"{o}}chl}(1994)}]{Blochl1994}%
  \BibitemOpen
  \bibfield  {author} {\bibinfo {author} {\bibfnamefont {P.~E.}\ \bibnamefont
  {Bl{\"{o}}chl}},\ }\href {\doibase 10.1103/PhysRevB.50.17953} {\bibfield
  {journal} {\bibinfo  {journal} {Phys. Rev. B}\ }\textbf {\bibinfo {volume}
  {50}},\ \bibinfo {pages} {17953} (\bibinfo {year} {1994})}\BibitemShut
  {NoStop}%
\end{thebibliography}
\end{document}